\documentclass[10pt, journal]{IEEEtran}
\usepackage[numbers]{natbib}
\usepackage[caption=false,font=normalsize,labelfont=sf,textfont=sf]{subfig}
\usepackage{soul}
\usepackage{hyperref}
\usepackage{amsmath,amssymb,amsfonts}
\usepackage{algorithmic}
\usepackage{graphicx}
\usepackage{textcomp}
\usepackage{xcolor}
\usepackage{multirow}
\def\BibTeX{{\rm B\kern-.05em{\sc i\kern-.025em b}\kern-.08em
    T\kern-.1667em\lower.7ex\hbox{E}\kern-.125emX}}
\begin{document}

\title{DeeP-TE: Data-enabled Predictive \\ Traffic Engineering}
\author{    
 ~Zhun Yin, Xiaotian Li, Lifan Mei,~\IEEEmembership{Member,~IEEE}, ~Yong Liu,~\IEEEmembership{Fellow,~IEEE}, ~Zhong-Ping Jiang, ~\IEEEmembership{Fellow,~IEEE} 

\thanks{\scriptsize{
*This work has been supported partially by NSF Grant CNS-2148309.

Zhun Yin, Xiaotian Li, Yong Liu and Zhong-ping Jiang are with the Department of Electrical and Computer Engineering at the New York University, Brooklyn, NY 11201, USA (emails: zy1652@nyu.edu, xl3399@nyu.edu, yongliu@nyu.edu, zjiang@nyu.edu).

Lifan Mei is with Xi'an Jiaotong-Liverpool University (email: Lifan.Mei@xjtlu.edu.cn).

(The first two authors contribute equally to the paper)
}

}}
\maketitle

\begin{abstract} 
Routing configurations of a network should constantly adapt to traffic variations to achieve good network performance. Adaptive routing faces two main challenges: 1) how to accurately measure/estimate time-varying traffic matrices? 2) how to control the network and application performance degradation  caused by frequent route changes? In this paper, we develop a novel data-enabled predictive traffic engineering (DeeP-TE) algorithm that minimizes the network congestion by gracefully adapting routing configurations over time. Our control algorithm can generate routing updates directly from the historical routing data and the corresponding link rate data, without direct traffic matrix measurement or  estimation. Numerical experiments on real network topologies with real traffic matrices demonstrate that the proposed DeeP-TE routing adaptation algorithm can achieve close-to-optimal control effectiveness with significantly lower routing variations than the baseline methods.
\end{abstract}


\section{Introduction}
Dynamic traffic demands pose significant challenges to network traffic engineering. In the traditional backbone networks, traffic oscillations can be ``absorbed" by network over-provisioning so that static or semi-static ``oblivious routing'' \cite{racke2009survey} can deliver good network performance. However, in contemporary communication networks, e.g., edge networks and intra-/inter- data center networks, not only the traffic demands are highly dynamic, but also the magnitude of the affordable/desirable over-provisioning is much lower than that of the traditional backbone networks. Good routing adaptation policies are much needed to efficiently utilize network resources to route dynamic traffic demands. With the recent fast development of SDN, which separates network operations into a control plane and a data plane, adaptive routing can be easily implemented at fine flow and temporal granularity.  However, adaptive routing still faces two main challenges: {\it 1) how to accurately measure/estimate time-varying traffic matrices? 2) how to control the network and application performance degradation  caused by frequent route changes?}    

One standard solution in the literature \cite{roughan2003traffic, otoshi2015traffic} is to periodically measure the current traffic matrix (TM), then re-calculate and implement optimal routing for the measured TM. Even though TM measurement has been extensively studied \cite{yassine2015software}, the accurate measurement of large-scale TMs is still not practical for SDN considering the limited number of entries in the Tenantry Content Addressable Memories (TCAM) of SDN switches \cite{wang2019joint}\cite{tian2018sdn}. Another solution is to measure link data rates, e.g., via the Simple Network Management Protocol (SNMP), and then infer the current TM. However, TM inference has to solve the Under-Determined Linear Inverse problem commonly faced by network tomography~\cite{vardi1996network}. Meanwhile, to control routing changes, routing configuration in one time interval has to consider the configurations in the previous and next time intervals. In \cite{otoshi2015traffic}, the authors proposed a Model Predictive Control (MPC) algorithm to address the problem of fast-varying routes in adaptive routing. In addition to minimizing the network delay, it explicitly adds route changes to its objective function so that routing adaptation is done ``gracefully''. However, their MPC algorithm requires the exact TM information in each control interval, which is hard to measure/estimate in real time.  

Inspired by the development of the behavioral system theory \cite{willems2005note, willems1991paradigms} and its recent related application Data-Enabled Predicitve Control (DeePC) \cite{berberich2022linear}, {\it we propose a DeeP-TE algorithm for adaptive routing that minimizes network congestion and controls route changes with only periodically sampled link data rates.} 
As a result, our proposed algorithm can adapt routes \emph{directly} based on historical link rate measurements without identifying the TMs in any intermediate steps. More specifically, link data rates are periodically measured at a predetermined sample interval, e.g., every 5 minutes, while periodical major routing updates are conducted at a longer control interval, e.g., every 1 hour. Without knowing the exact traffic matrix, following the behavioral approach to system theory, we develop a data-driven model that predicts the impact of potential routing changes on link data rates based on historical link rate measurements under different routing configurations. For accurate predictions, the historical data samples should be collected from a rich set of ``persistently exciting'' routing configurations. We theoretically study the minimum number of data samples needed for full characterization of routing impacts. We further reduce the number of data samples by focusing only on dominating elephant flows and aggregating routing configurations using network topology information. To adapt to traffic variations within a control interval and generate rich data samples, we introduce heuristic routing perturbations for a small fraction of demands at each sample interval. Numerical experiments on real network topologies and TMs demonstrate that the proposed DeeP-TE routing adaptation algorithm can achieve close-to-optimal control effectiveness with significantly lower routing variations than the baseline methods.

The rest of the paper is organized as follows: The related work is surveyed in Section~\ref{sec:related}. We present a preliminary of data-enabled predictive control in Section~\ref{sec:prelim}. The adaptive routing problem is defined in Section~\ref{sec:routing}. The DeePC-based routing adaptation is formulated in Section~\ref{sec:MPC}. The traffic dynamic model and system order reduction are presented in Section~\ref{sec:reduction}. The complete DeeP-TE algorithm is developed in Section~\ref{sec:algorithm}. Experimental results with real network topologies and TMs are presented in Section~\ref{sec:experiments}. The paper is concluded with future work in Section~\ref{sec:conclusion}.  

\subsection{Related Work}
\label{sec:related}
Model predictive control, also known as receding horizon control (RHC), for traffic engineering was originally studied in \cite{otoshi2015traffic} to control routing changes. The authors further proposed a robust variant in \cite{otoshi2015traffic} and a hierarchical variant in \cite{otoshi2018hierarchical}. Most recently, two RHC-based TE algorithms have been proposed in \cite{zheng2021online} to tradeoff TE goal with route changes. Another direction that can jointly optimize TE goal and route changes is reinforcement learning. In \cite{ye2022mitigating}, the authors train an actor network to select the ``critical'' flows and only change the routes for them to mitigate the route changes. However, all the above methods require the exact knowledge of the TMs. 

Among the TE algorithms that do not require the measurement of TMs, ``tomogravity MPLS'' and ``tomogravity HuerOSPF'' are perhaps the most famous two \cite{roughan2003traffic}. They both use the TMs estimated by the tomogravity method \cite{zhang2003fast} to do traffic engineering. However, as pointed out in \cite{soule2007estimating}, some carriers indicate that they will only consider using TMs for TE if the average estimation error is below $10\%$, which is hard to achieve for most network tomography methods. Many existing methods for TM estimation utilize additional information, such as past complete TMs and/or TM distributions, to approximately solve the UDLI problem~\cite{zhao2019spatiotemporal}. The authors of \cite{soule2007estimating} propose to use deliberate route changes as perturbations to estimate the TMs. We also induce some deliberate route perturbations, but we do \textit{not} estimate the TMs. Instead, our algorithm generates routing configurations directly from the perturbation route data and the corresponding link load data. Our algorithm accounts for traffic variations within a control interval, which is impossible for the traditional periodic TM estimations  followed by routing updates. 

Machine learning, in particular, Reinforcement Learning and Deep Reinforcement Learning (DRL) have recently found their applications in intra- and inter-domain routing, e.g.,~\cite{DRL-Routing1,DRL-Routing2,23teal}. A naive solution is to train a popular DRL agent that treats the network as a blackbox and uses generic DNNs as its critic and actor, which will be ``magically" tailored through training to generate good routing adaptation policies. Due to the complexity of network and traffic dynamics, large training samples and excessively long training time are expected, and the DRL agent will most likely be topology-specific and does not work well when there are major changes in traffic demands.  On the other hand, our DeeP-TE algorithm leverages the well-defined network routing model to achieve much higher data sample efficiency with theoretic guarantees. Heavy-tailed traffic distribution and known network topology information can be further exploited to significantly reduce the number of data samples needed to drive MPC calculation.

\section{Preliminaries}
\label{sec:prelim}
We first describe the classical model predictive control, then introduce the recently developed data-driven model predictive control.

\textbf{Notations.} We use $\mathbb{R}$ and $\mathbb{N}$ to represent the set of real numbers and the set of natural numbers, respectively. We use $\mathbb{R}^n$ to represent the $n$-dimensional Euclidean space.
\subsection{Model Predictive Control (MPC)} \label{subsec:MPC}
Model predictive control is a discrete-time control strategy that generates control decisions by first computing a trajectory of future control inputs with the best predicted cumulative future outcomes for a given prediction horizon at each time step, and then only implementing the nearest future control in the computed  trajectory, in which the predictions are typically generated by feeding the measured outputs and the predicted control inputs into a predetermined \textbf{prediction model}.

For example, in this work, we mainly employ the MPC controller designed for linear time-invariant systems with the following minimal representation:
\begin{subequations} \label{LTI-sys}
\begin{alignat}{2}
    x(k+1)=Ax(k)+Bu(k), \\
    y(k)=Cx(k)+Du(k)
\end{alignat}
\end{subequations}
where $x(k)\in\mathbb{R}^{n}$, $u(k)\in\mathbb{R}^{m}$ and $y(k)\in\mathbb{R}^p$ represent the state variables, control inputs, and system outputs, respectively, and matrices $\{A,B,C,D\}$ are the \textbf{linear system model}. Notably, in this case, since we can predict the future state variables $x(k+H)$ for any prediction horizon $H\in\mathbb{N}$ using the measured current state variables $x(k)$ and the predicted future control inputs $\{u(k+h-1)\}_{h=1}^{H}$, the system model $\{A,B,C,D\}$ can also serve as the prediction model.

At the time step $k$, the MPC controller with a prediction horizon $H$ for the above system Eq.~\eqref{LTI-sys} solves the following optimization problem:
\begin{subequations} \label{MPC}
\begin{alignat}{2}
& \underset{u,x,y}{\textbf{minimize}} & & \; J\left(y({[k+1,k+H]}), u({[k,k+H-1]})\right) \label{MPC_obj} \\
& \text{s.t.} & & \text{For $h=1,\dots,H$:} \nonumber \\ 
& & & \begin{cases}
    x(k+h)=Ax(k+h-1)+Bu(k+h-1), \\
    y(k+h)=Cx(k+h)+Du(k+h), 
\end{cases} \label{MPC:sub2} \\
& & & x(k) = x_0, \label{MPC:sub3} \\
& & & u([k,k+H-1]) \in \mathcal{U}, \label{MPC:sub4} \\
& & & y([k+1,k+H]) \in \mathcal{Y}, \label{MPC:sub5}
\end{alignat}
\end{subequations}
where $x_0$ is the measured or estimated system state at the current time step $k$, $J(\cdot)$ is the cost function, $\mathcal{U}$ and $\mathcal{Y}$ represent the constraints for the control inputs and outputs, respectively. We denote the solution of the above optimization problem Eq.~\eqref{MPC} as $u^*({[k,k+H-1]})$, and we only implement the control input $u^*(k)$ at time step $k$. For the next time step, the MPC controller repeats the whole process by letting $k:=k+1$. This is to minimize the negative impact of unavoidable prediction errors, especially at farther away future time steps. 

\subsection{Data-Enabled Predictive Control (DeePC)} \label{subsec:DeePC}

The main difficulty with the MPC algorithm in the previous section is that system model $\{A,B,C,D\}$ may be unavailable, or inaccurate, which will affect the control effectiveness. Additionally, the inital system state $x(k)$ also may not be accurately measured/estimated. Recently, an algorithm called Data-Enabled Predictive Control (DeePC) was proposed in \cite{coulson2019data} and it can generate the control inputs that are equivalent to those generated by the MPC controller for the linear time-invariant system directly from the sampled input/output data without identifying the system model, nor explicitly estimating the initial state.  Before formally introducing the algorithm, we first introduce the following notations, definitions, and propositions. 
Given a sequence of vectors, $\{z(a), \dots, z(b)\}$, 
\begin{equation}
    z_{[a,b]} := [z^T(a),z^T(a+1),\dots,z^T(b)]^T. \nonumber
\end{equation}
We define the Hankle matrix of the vector sequence as:
\begin{equation}
    \mathcal{H}_l(z_{[a,b]}) := 
    \begin{bmatrix} 
    z(a), \cdots, z(b-l+1) \\ 
    \vdots\;\; \ddots \;\;\vdots \\
    z(a+l-1), \cdots, z(b)
    \end{bmatrix}, \nonumber
\end{equation}
and we denote $H^{(z)}_{a,b}:=\mathcal{H}_1(z_{[a,b]})$

\noindent\textbf{Definition 1} (\cite{willems2005note})\textbf{.} \textit{For the system in Eq.~\eqref{LTI-sys}, we say a trajectory of control input $u_{[0,L-1]}$, $u(k)\in\mathbb{R}^{m}$, is persistently exciting of order $l\le L$, if the following equation holds:
\begin{equation}
    rank\left(\mathcal{H}_{l}(u_{[0,L-1]})\right) = ml.
\end{equation}
}
\\
\noindent\textbf{Theorem 1} (\cite{willems2005note})\textbf{.} \textit{If the system in Eq.~\eqref{LTI-sys} is controllable, given a control trajectory $u_{[0,L-1]}$ that is persistently exciting of order $(l+n)$, $y_{[0,L-1]}$ is the corresponding system output trajectory, then any other control/output trajectory pairs of length $l \le L$ must be in the space spanned by the columns of the Hankle matrix of the given control/output pair: $\mathcal{H}_{l}\left([u^T_{[0,L-1]},y^T_{[0,L-1]}]^T\right)$. 
}
\\

Intuitively, persistently exciting control sequence $u_{[0,L-1]}$ provides rich enough stimuli to the system so that its dynamics can be completely characterized by the output 
data samples $y_{[0,L-1]}$, without using the linear system model.

To obtain this ``data-driven" system model, we can collect control/output data samples offline. According to \cite{coulson2019data}, to satisfy the conditions of Theorem 1, the number of control/output data samples $L$ should be no less than $(m+1)(l+n)-1$. Let $[u^T_{data}, y^T_{data}]^T$ be a historical system trajectory such that $u^T_{data}$ is persistently exciting of order $(l+n)$. For online operation, let $[u_{ini}^T, y_{ini}^T]^T$ be the measured control/output trajectory in the previous $l_{ini} < l$ steps, $[u_{pred}^T, y_{pred}^T]^T$ be the predicted control/output trajectory in the next $l_{pred}=l-l_{ini}$ steps. According to Lemma 1 in \cite{Markovsky2008DatadrivenSA} and Theorem 1, we will have:
\begin{equation} \label{data-driven}
    \begin{bmatrix}
        U_p \\
        Y_p \\
        U_f \\
        Y_f \\
    \end{bmatrix} g = \begin{bmatrix}
        u_{ini} \\
        y_{ini} \\
        u_{pred} \\
        y_{pred} \\
    \end{bmatrix},
\end{equation}
where vector $g\in\mathbb{R}^{L-l+1}$is called the ``decision variables'', 
\begin{equation}
    \begin{bmatrix}
        U_p \\
        U_f
    \end{bmatrix} = \mathcal{H}_{l}(u_{data}), \;
    \begin{bmatrix}
        Y_p \\
        Y_f
    \end{bmatrix} = \mathcal{H}_{l}(y_{data}), 
\end{equation}
i.e., $U_p$ and $U_f$ are the first $l_{ini}$ and the last $l-l_{ini}$ row blocks of the Hankle matrix of the historical control trajectory, $Y_p$ and $Y_f$ form a similar partition of the Hankle matrix of the historical output. Consequently, the MPC controller in Eq.~\eqref{MPC} can be re-written as the following data-driven expression by replacing the system model Eq.~\eqref{MPC:sub2} and initial condition Eq.~\eqref{MPC:sub3} by Eq.~\eqref{data-driven}:
\begin{subequations} \label{data-driven MPC}
\begin{alignat}{2}
& \underset{u_{pred},y_{pred},g}{\textbf{minimize}} & & \; J\left(y_{pred}, u_{pred}\right) \label{data-driven MPC_obj} \\
& \text{s.t.} & & \begin{bmatrix}
        U_p \\
        Y_p \\
        U_f \\
        Y_f \\
    \end{bmatrix} g = \begin{bmatrix}
        u_{ini} \\
        y_{ini} \\
        u_{pred} \\
        y_{pred} \\
    \end{bmatrix}, \label{data-driven MPC:sub2} \\
& & & u_{pred} \in \mathcal{U}, \label{data-driven MPC:sub4} \\
& & & y_{pred} \in \mathcal{Y}. \label{data-driven MPC:sub5} 
\end{alignat}
\end{subequations}
If $l_{ini}$ is long enough, the initial system state $x_0$ in Eq.~\eqref{MPC} can be uniquely obtained using recent measurements $u_{ini}$ and $y_{ini}$, and the data-enabled predictive control solution is identical to the model-based MPC solution.

\section{Problem Formulation}
\label{sec:routing}
\subsection{Network Model}
Without loss of generality, we model a communication network as a directed graph $G = (V,E)$, where $V$ represents the set of all nodes and $E$ represents the set of all links. We denote the number of links as $n_l:=|E|$ and the capacity of the link $l_i\in E$ as $C_{l_i}$. Denote the capacity vector of all links as $C = [C_{l_1}, \dots, C_{l_{n_l}}]^T$. The maximal number of source-destination demand pairs 
is $n_w:=|V|(|V|-1)$. For each demand $d$, there is a set of $n_{p,d}$ pre-configured candidate routing paths. The total number of routing paths in the network is  $n_p:=\sum_{d=1}^{n_w} n_{p,d}$. For the readers' convenience, we list all the notations in the Table~\ref{tab:notations} which classifies the notations according to the sections in which they are defined.

\begin{table}[h!] \label{tab:notations}
\caption{Table of Notattions}
\small
\begin{center}
\begin{tabular}{|c|c|}
\hline
Notations & Meanings \\
\hline
$z_{[a,b]}$ & $[z^T(a),z^T(a+1),\dots,z^T(b)]^T$ \\
$\mathcal{H}_{l}(z_{[a,b]})$ & Hankle matrix of $z_{[a,b]}$ with $l$ rows\\
$H_{a,b}^{(z)}$ & $\mathcal{H}_{1}(z_{[a,b]})$ \\
$g$ & vector of decision variables \\
\hline
$G$ & commuunication network \\
$V$ & set of all nodes \\
$E$ & set of all links \\
$d$ & index for demands \\
$l_i$ & index for links \\
$C_{l_i}$ & link capacity of link $l_i$ \\
$t$ & time \\
$n_l$ & number of links \\
$n_w$ & number of demands \\
$n_p$ & total number of paths \\
$n_{p,d}$ & number of path for demand $d$ \\
\hline
$\tau$ & sample interval \\
$\Tilde{w}$ & vector of continuous demands \\
$w$ & vector of discretized demands \\
$s$ & index for sample intervals \\
\hline
$y$ & vector of link loads \\
$P_d$ & constant routing matrix for demand $d$ \\
$r$ & vector of all routing configurations \\
$r_{[d]}$ & vector of routing configurations for demand $d$ \\
$r_{[d],j}$ & fraction of demand $d$ routed to its $j$-th path \\
\hline
$\tau_c$ & control interval \\
$N$ & \#sample intervals within each control interval \\
$\Bar{w}$ & piecewise constant demands \\
$k$ & index for control intervals \\
$f$ & function of delay \\
\hline
$\Hat{w}$ & vector of predicted piece-wise constant demands \\
$X$ & constant matrix for the simple prediction model \\
$H$ & maximumal prediction horizon minus one \\
$L$ & maximumal past horizon\\
$y^{p,h}(Nk)$ & vector of link loads (use $r(N(k+h))$ for $\Bar{w}(k-p)$) \\
$g^{p,h}(Nk)$ & vector of decision variables for $y^{p,h}(Nk)$ \\
\hline
$n_{\phi}$ & number of basis functions \\
$\phi_i (\cdot;k)$ & $i$-th basis function within the $k$-th control interval \\
$\Bar{\rho}_i$ & vector of weights for the $i$-th basis function for $\Tilde{w}$ \\
\hline
$b_d$ & index for elephant flows \\
$n'_w$ & number of elephant flows \\
$r'$ & vector of routing configurations of elephant flows \\
$n'_p$ & total number of paths of elephant flows \\
$n'_{p,b_d}$ & number of path for elephant flow $b_d$ \\
$c_{l_i}$ & the $i$-th link load consisted of mice flows \\
$c$ & vector of the link loads consisted of mice flows \\
$\Bar{c}$ & mean value of the link load consisted of mice flows \\
$\Bar{c}_i$ & vector of weights for the $i$-th basis function for $c$\\
\hline
$r^{agg}_{l_i}$ & vector of the $i$-th link's aggreated routing fractions \\
$r^{agg}$ & matrix consisted of the aggreated routing fractions \\
$M_{l_i}$ & aggreation matrix for the $i$-th link \\
$M^{agg}$ & aggreation matrix \\
$e_{l_i}$ & indicating vector for the $i$-th link \\
\hline
$J_h$ & cost function of horizon $h$ \\ 
$\Delta r$ & routing configuration differences \\
$\alpha_1$ & hyper-parameter for the rout changes cost \\ 
$\alpha_2$ & hyper-parameter for the decision variable regulation \\
$0_{n}$ & all zero vector of dimension $n$ \\
$1_{n}$ & all one vector of dimension $n$ \\
$S$ & matrix representing the basis functions \\
\hline
\end{tabular}
\label{table:notations}
\end{center}
\end{table}

\subsection{Multi-Timescale Measurement and Control} \label{Multi-time Scale}
We consider time-varying traffic demands. In a real-world setting, the traffic demands are continuous, and we denote the continuous demands as $\Tilde{w}(t): \mathbb{R}_+\rightarrow\mathbb{R}_+^{n_w}$. The network is measured/monitored periodically at discrete time instants with \textbf{sample interval} of $\tau$. We define the discretized traffic demands as $w(s) := \Tilde{w}(s\tau)$. While link rates can be measured periodically, e.g. using SNMP every 5 minutes, it is unfortunately impossible to completely recover the traffic demands just based on link rate measurements,  due to the Under-Determined Linear Inverse (UDLI) problem commonly faced by network tomography~\cite{vardi1996network}. The per-flow measurement capability of SDN can be used to monitor traffic flow between source-destination pairs. However, it is expensive to monitor all flows at small sample intervals. {\it One main challenge we address in this paper is to optimally adapt network routing without knowing the dynamic traffic matrix $w(s)$.} 

The link loads at time step $s$ is denoted by $y(s)=[y_{l_1}(s),\dots,y_{l_i}(s), \dots,y_{l_{n_l}}(s)]^T$, where $y_{l_i}(s)$ represents the load of link $l_i \in E$, and can be expressed by the following link-path formula:
\begin{equation}
y(s)=\sum_{d=1}^{n_w} P_d w_d(s)r_{[d]}(s)=D(w(s))r(s) \label{sys}, 
\end{equation}
where $P_d \in \mathbb{R}^{n_l\times n_{p,d}}$ is constant routing matrix with each column representing a path (the $i$-th element of the column equals to 1 if the $i$-th link is in the corresponding path, 0 otherwise) for demand $d$, and $r(s):=[r^T_{[1]}(s), \cdots, r^T_{[n_w]}(s)]^T\in\mathbb{R}^{n_p}$ denotes the routing configurations at time step $s$, in which $r_{[d]}(s)\in[0,1]^{n_{p,d}}$ and $r_{[d],j}(s)$ represents the fraction of  demand $d$ routed to its $j$-th candidate path at time step $s$. Note that $D(w(s))=[w_1 (s)P_1, \cdots, w_{n_w}(s)P_{n_w}]$.

 To avoid frequent route changes, we only adapt the routing configurations at an interval longer than the sample interval. Similar to the ``snapshot'' defined in \cite{soule2007estimating}, we define the interval between two adjacent routing updates as the \textbf{control interval} $\tau_c$. We assume each control interval covers multiple sample intervals, i.e., $\tau_c=N\tau$. For example, if the routing is updated once every hour based on SNMP measurements every 5 minutes, then we have $N=12$. 
 
One naive adaptive routing solution is to run independent routing optimization at the beginning of each control interval by assuming piece-wise constant traffic matrix, i.e., $\Tilde{w}(t) \approx \Bar{w}(k)$, $\forall t \in [k\tau_c, (k+1) \tau_c)$, where $\Bar{w}(k)$ stands for the estimated average traffic matrix in the $k$-th control interval. And the routing is static until the next control interval, i.e., $r(s)=r(Nk)$, $\forall s \in (Nk, N(k+1))$. This approach does not consider the traffic variations within each control interval, and there is no control on routing configuration changes between adjacent control intervals. Major routing configuration changes are expensive to 
implement. It also takes more time for the network to stabilize, introducing significant interruptions to the network and application performances. To address these issues, we employ more refined dynamic traffic model and study the routing adaptation problem as an  optimal control problem to explicitly control the routing changes. 

\subsection{Optimal Control for Traffic Engineering}

Given a session of $M$ control intervals, adaptive routing can be formulated as an optimal control problem.
\begin{subequations} \label{optimal_control}
\begin{alignat}{2}
& \underset{r(s)}{\textbf{min}} & & \; \sum_{s=1}^{MN}  \sum_{i=1}^{n_l} f\left(\frac {y_{l_i}(s)} {C_{l_i}}\right) 
+ \alpha_1 \sum_{k=2}^M \|r(kN)-r((k-1)N)\|_1 \label{oc:sub0}
\\
& \text{s.t.} \; & &  \text{Eq.~\eqref{sys}}, \sum_{j=1}^{n_{p,d}} {r}_{[d],j} (s) = 1, \forall d=1,\dots,n_w,\label{oc:sub1} \\
&&& \quad 0\preceq r(s) \preceq 1, r(s)=r(Nk), \forall s \in [Nk, N(k+1)),
\end{alignat}
\end{subequations}
where the first term in ~\eqref{oc:sub0} is the total link delay as a function of link utilization, and the second term measures the routing changes between two adjacent control intervals, $\alpha_1$ controls the trade-off between the two objectives.  With known traffic matrices  in Eq.~\eqref{sys} in all control intervals, the problem can be solved as a gigantic linear/convex optimization problem to obtain piece-wise constant routing adaptation trajectory $\{r(s)\}$ to achieve the desirable trade-off between network delay performance and routing update cost in the face of dynamic traffic demand variations.

\section{Data-enabled Predictive Control based Routing Adaptation}
\label{sec:MPC}

For online routing adaption, not only the future traffic matrices are not available, the past and current traffic matrices may not be directly measurable. We resort to the recent developments in Data-enabled Predictive Control \cite{coulson2019data} to solve the previous adaptive routing problem without knowing dynamic traffic matrices. 

For traffic engineering, the control input is the routing configuration $r(s)$ and the output is its corresponding link loads $y(s)$. The input/output relationship defined in Eq.~\eqref{sys} is stateless ($n=0$), so that we do not need to estimate $x_0$, and therefore we can set $l_{ini}=0$. According to Theorem 1, the longer the prediction length $l$ is, the more data we need to sample. In fact, there is no need to use any $l>1$ for stateless systems  \cite{berberich2020trajectory}. As a result, the minimal data points needed will be $(m+1)\times(1+0)-1=m$, which corresponds to the total number of independent control variables. For example, $m=n_p-n_w$ for the system Eq.~\eqref{sys} if the $n_w$ constraints Eq.~\eqref{oc:sub1} are considered.
We can simplify Eq.~\eqref{data-driven MPC:sub2} into:
\begin{equation} \label{data-driven TE}
\begin{aligned}
   & \begin{bmatrix}
        H_{0,m-1}^{(r)}(\bar w(k)) \\
        H_{0,m-1}^{(y)}(\bar w(k)) \\
    \end{bmatrix} g = \begin{bmatrix}
        r(k) \\
        y(k) \\
    \end{bmatrix}, \\
\end{aligned}
\end{equation}
where $k$ is the current control interval, $H_{0,m-1}^{(r)}(\bar w(k))$ represents $m$ past routing configurations applied to the constant TM $\bar w(k)$, and $H_{0,m-1}^{(y)}(\bar w(k))$ represents the  corresponding sampled link load data. For any chosen $r(k)$, the link loads $y(k)$ can be predicted by solving Eq.~\eqref{data-driven TE} using the past $m$ routing-link load pairs, without knowing the TM $\bar w(k)$.  

Following the MPC framework, to calculate $r(k)$, one needs to predict the future link loads $y(k+h)$ under future routing configuration $r(k+h)$ without knowing the future traffic matrix $\bar w(k+h)$, nor any routing-load pair samples obtained from future TM $\bar w(k+h)$. We use a simple offline trained TM prediction model to get around this obstacle.  


In particular, similar to \cite{otoshi2018hierarchical}, we use a linear model to predict the mean value of the traffic demands at the next $H$ ($H$ is some constant positive integer) control intervals as follows:
\begin{equation}
\begin{aligned}
& \begin{bmatrix} \Hat{w}(k+H-1), \cdots, \Hat{w}(k)\end{bmatrix}= \\
& \begin{bmatrix} \Bar{w}(k-L), \cdots, \Bar{w}(k-1)\end{bmatrix} X,
\end{aligned}
\end{equation}
where $\Hat{w}(\cdot)$ represents the predicted value of $\Bar{w}(\cdot)$. And we denote $X$ as follows:
\begin{equation}
X = \begin{bmatrix} 
    X_{L,H-1}, \cdots, X_{L,0} \\ 
    \vdots\;\; \ddots \;\;\vdots \\
    X_{1,H-1}, \cdots, X_{1,0}
    \end{bmatrix}
\end{equation}
The model's parameter $X$ is trained from the historical TM data set. 
With linear TM prediction, based on~\eqref{sys}, we can predict the future link load as 
\begin{equation}
\begin{aligned}
y\left(N(k+h)\right) &=  D(\Hat{w}(k+h)) r(N(k+h))\\ 
          & =  D\left(\sum_{p=1}^{L} X_{p,h}\Bar{w}(k-p)\right) r(N(k+h)) \\
 &=\left[\sum_{p=1}^{L} X_{p,h}D(\Bar{w}(k-p))\right]r(N(k+h)) \\
  &=\sum_{p=1}^{L} X_{p,h} y^{p,h}(Nk), 
 \label{yp=Br11}
\end{aligned}
\end{equation}
where $y^{p,h}(Nk)$ is the ``dummy" link load vector if routing configuration $r(N(k+h))$ was applied to the traffic matrix $\bar w(k-p)$, which can be solved using equations driven by data samples collected from the control interval $k-p$, similar to \eqref{data-driven TE}:  
\begin{equation} \label{data-driven TE1}
\begin{aligned}
   & \begin{bmatrix}
        H_{0,m-1}^{(r)}(\bar w(k-p)) \\
        H_{0,m-1}^{(y)}(\bar w(k-p)) \\
    \end{bmatrix} g^{p,h}(Nk) = \begin{bmatrix}
        r(N(k+h)) \\
        y^{p,h}(Nk)\\
    \end{bmatrix}, \\
\end{aligned}
\end{equation}
 where $H_{0,m-1}^{(r)}(\bar w(k-p))$ are past routing configurations applied in control interval $k-p$, and 
 $H_{0,m-1}^{(y)}(\bar w(k-p))$ are corresponding link load vectors.  


\section{Traffic Dynamics and System Order Reduction} 
\label{sec:reduction}
The piecewise constant routing control in~\eqref{optimal_control} keeps routing unchanged within each control interval, therefore ignoring traffic variations within a control interval. Meanwhile, the DeeP-TE calculation in~\eqref{data-driven TE1} needs a sufficient number of historical routing-link load samples. In this section, we describe in detail how to handle traffic variations within each control interval and how to reduce the number of data samples needed through system order reduction. 

\subsection{Traffic Dynamics within Control Interval}
\label{sec:5.a}
Within the $k$-th control interval, we assume that the traffic demands can be modeled as: 
\begin{equation} 
\begin{aligned}
\Tilde{w}(t) \approx & \; \Bar{w}(k)+\sum_{i=1}^{n_{\phi}} \Bar{\rho}_i(k) \phi_i(t;k) \\
& \; \text{for} \; kN\tau \le t < (k+1)N\tau
\end{aligned}
\end{equation}
where $\Bar{w}(k)\in\mathbb{R}^{n_w}$ represents the mean value of the traffic demands in the $k$-th control interval, each $\phi_i(t;k)$ is a known $\mathbb{R}$-valued basis function that models the cyclo-stationarity feature of the traffic demands which satisfies the following:
\begin{equation}
\int_{kN \tau}^{(k+1)N \tau} \phi_i(t;k) dt = 0, \; k\in\mathbb{N},
\end{equation}
and the vector $\Bar{\rho}_i(k)\in\mathbb{R}^{n_w}$ are unknown coefficients for the $i$-th basis function, and $n_{\phi}\in\mathbb{N}$ represents the total number of basis functions.
Obliviously, we have:
\begin{equation} \label{disc demand model}
\begin{aligned}
w(s) \approx & \; \Bar{w}(k)+\sum_{i=1}^{n_{\phi}} \Bar{\rho}_i(k) \phi_i(s\tau;k) \\
& \; \text{for} \; Nk \le s \le N(k+1)
\end{aligned}
\end{equation}

By replacing $w(s)$ with Eq.~\eqref{disc demand model} in Eq.~\eqref{sys}, we can get the following system model:
\begin{equation} \label{piece-wise sys}
\begin{aligned}
& y(s) =D\left(\Bar{w}(k)+\sum_{i=1}^{n_{\phi}} \Bar{\rho}_i(k) \phi_i(s\tau;k)\right)r(s), \\
& = D\left(\Bar{w}(k)\right)r(s) + \sum_{i=1}^{n_{\phi}} \phi_i(s\tau;k) D\left(\Bar{\rho}_i(k)\right)r(s), \\
& = \left[D\left(\Bar{w}(k)\right), D\left(\Bar{\rho}_1(k)\right), \dots, D\left(\Bar{\rho}_{n_\phi}(k)\right)\right] \begin{bmatrix}
    r(s) \\
    \phi_1(s\tau;k)r(s) \\
    \vdots \\
    \phi_{n_\phi}(s\tau;k)r(s)
\end{bmatrix}\\
& \; \text{for} \; Nk \le s \le N(k+1), \; k\in\mathbb{N}.
\end{aligned}
\end{equation}

Now we can notice that the link-path formulation can be expressed as a linear time-invariant system in the form of Eq.~\eqref{LTI-sys} with $A,B,C$ matrices all being zero matrices and the $D\in\mathbb{R}^{n_l \times (1+n_\phi)n_p}$ matrix being constant within each control interval. If we denote the last vector in the Eq.~\eqref{piece-wise sys} as control input $u$ and assume that the data are sampled from the interval $[Nk+a, Nk+b]$ with $0\le a< b\le N$, the Hankle matrix satisfying the condition in Theorem 1 with $l=1$ is:
\begin{equation}
\begin{aligned}
& H^{(r)}_{Nk+a,Nk+b} = \\ 
& \begin{bmatrix}
    \scriptstyle{r(Nk+a)} & \dots & \scriptstyle{r(Nk+b)}, \\
    \scriptstyle{\phi_1((Nk+a)\tau;k)r(Nk+a)} & \dots & \scriptstyle{\phi_1((Nk+b)\tau;k)r(Nk+b)}\\
    \vdots & \dots & \vdots  \\
    \scriptstyle{\phi_{n_\phi}((Nk+a)\tau;k)r(Nk+a)} & \dots & \scriptstyle{\phi_1((Nk+b)\tau;k)r(Nk+b)}
\end{bmatrix}
\end{aligned}
\end{equation}
To make this Hankle matrix a full row rank matrix, the block consisting of the first $n_p$ rows of the Hankle matrix must be full row rank; this means we should have at least $n_p$ different routes. However, the rest block consisting of the last $n_p \times n_{\phi}$ rows of the Hankle matrix can be made full row rank by sampling the same route at different time steps. That is, we do not need more routing samples to work with the basis functions.

Nevertheless, to implement the DeePC algorithm for the derived system Eq.~\eqref{piece-wise sys}, the number of required different routing configurations is still very large even for a relatively small network.
For the Geant network with 22 nodes and 36 links, if we select the top four shortest paths for each OD flow to split, then $n_p=1484$, which is also the minimal number of required different routing configurations for each control interval.
Since we would like to reduce the route changes, using such large number of different routes for one control interval are impractical. To enhance the data efficiency, we propose two methods to reduce the input dimension of the system Eq.~\eqref{piece-wise sys}.

\subsection{Utilize the Heavy-tail Property of Traffic Matrices}
\label{sec:5.b}

The heavy-tail property of TM has been observed in \cite{soule2007estimating}\cite{xie2023deep}.
A small number of flows carry the most Internet traffic, while the remainder of flows are large in numbers but account for only a very small portion.
Similar to other works, we call the large flows ``elephant'' flows and the rest the ``mouse'' flows.
Generally, if we have $n'_w$ elephant flows in total, we denote the set of the indices of elephant flows as $\{b_1,b_2,\dots,b_{n'_w}\}\subseteq\{1,2,\dots,n_w\}$ and the corresponding routing vector at time step $s$ is defined by $r'(s):=[r^T_{[b_1]}(s), \cdots, r^T_{[b_{n'_w}]}(s)]^T$. 
And we define $n'_p:=\sum_{d=1}^{n'_w} n_{p,b_d}$. 




To reduce the order of the system Eq.~\eqref{piece-wise sys}, the routing configurations of mice flows are calculated based on historical data, e.g., routing configurations during the same hour of the previous day, and remain unchanged for the all control intervals.
And the link load part accounting for the mice flows is denoted by $c_{l_i}(s)$ for each $l_i \in E$. And we define the vector $c(s)=[c_{l_1}(s),\dots,c_{l_{n_l}}(s)]^T$. 
Since the routes are constant and according to the model Eq.~\eqref{piece-wise sys}, $c(s)$ can be expressed by the following equation that is similar to Eq.~\eqref{disc demand model}:
\begin{equation} \label{disc mice demand model}
\begin{aligned}
c(s) \approx & \; \Bar{c}(k)+\sum_{i=1}^{n_{\phi}} \Bar{c}_i(k) \phi_i(s\tau;k) \\
& \; \text{for} \; Nk \le s \le N(k+1)
\end{aligned}
\end{equation}
where $\Bar{c}(k)\in\mathbb{R}^{n_l}$ represents the mean value of the link load part accounting for the mice flows in the $k$-th control interval, and $\Bar{c}_i(k)\in\mathbb{R}^{n_l}$ represents the vector of unknown coefficients for the basis function $\phi_i$ for each link in the $k$-th control interval. The reduced order system can be rewritten as follows:
\begin{equation} \label{reduced order sys 1}
\begin{aligned}
y(s)= & \left[D'\left(\Bar{w}(k)\right), D'\left(\Bar{\rho}_1(k)\right), \dots, D'\left(\Bar{\rho}_{n_\phi}(k)\right)\right] \\ 
& \cdot \begin{bmatrix}
    r'(s) \\
    \phi_1(s\tau;k)r'(s) \\
    \vdots \\
    \phi_{n_\phi}(s\tau;k)r'(s)
\end{bmatrix} + \Bar{c}(k)+\sum_{i=1}^{n_{\phi}} \Bar{c}_i(k) \phi_i(s\tau;k)\\
& \; \text{for} \; Nk \le s \le N(k+1), \; k\in\mathbb{N}. 
\end{aligned}
\end{equation}
where $D'(\bar{w}(k)):=[\bar{w}_{b_1}(k)P_{b_1}, \cdots, \bar{w}_{b_{n'_w}}(k)P_{b_{n'_w}}]$ and $D'(\Bar{\rho}_i(k)):=[\bar{\rho}_{i,b_1}(k)P_{b_1}, \cdots, \bar{\rho}_{i,b_{n'_w}}(k)P_{b_{n'_w}}]$.

\subsection{Utilize the Topology Information}
\label{sec:system-order-c}

Though by only controlling the elephant flows we can reduce the model order by about two-thirds, the minimal perturbations needed are still too many;
e.g., for $n_\phi=2$, over one hundred perturbations are still needed for the Abilene topology even though we only control the top ten elephant flows and each elephant flow can split among at most four paths. 
Inspired by the heuristic in \cite{nucci2004design}, we can treat each link instead of the whole network as our system. The model for link $l_i$ is:
\begin{equation}
y_{l_i}(s) = D'(w(s))_{l_i,*}r'(s) + c_{l_i}(s), \label{piece-wise constant link sys}
\end{equation}
where $D'(w(s))_{l_i,*}$ denotes the row of the matrix $D'(w(s))$ which corresponds to link $l_i$.

We notice that for each link $l_i$, $D'(w(s))_{l_i,*}$ contains many repeated elements and zeros. Thus, we propose to use the ``aggregated" routing fractions $r^{agg}_{l_i}\in\mathbb{R}_{+}^{n'_w}$ as the input, the $d$-th element of which represents the total fraction of the $b_d$-th OD flow allocated to the link $l_i$, which is the summation of the routing fractions on all its candidate paths that traverses link $l_i$.  And we define matrix $r^{agg}:=[r^{agg}_{l_i},\cdots,r^{agg}_{l_{n_l}}]$.


The transformation from $r'$ to $r^{agg}_{l_i}$ is represented by an aggregation matrix $M_{l_i}\in\mathbb{R}^{n'_w\times n'_p}$ where $M_{l_i,d,p}=1$ if the $b_d$-th OD flow can be split to the $p$-th path and this $p$-th path contains link $l_i$, otherwise $M_{l_i,d,p}=0$. Then, we have:
\begin{equation} \label{M^agg}
    r^{agg}=    
    \begin{bmatrix}
        M_{l_1} \\
        \vdots \\
        M_{l_{|E|}}
    \end{bmatrix} r' =: M^{agg} r'
\end{equation} 
Note that in order to get $M_{l_i}$ we need to know the topology information of the network.

Then we can have the following reduced order system:
\begin{equation}
y_{l_i}(s)=[w_{b_1}(s), \cdots, \bar{w}_{b_{n'_w}}(t)]r^{agg}_{l_i}(s) +  c_{l_i}(s), \label{reduced order sys 2}
\end{equation}
By regarding $c(s)$ as part of the system parameters, we have:
\begin{equation}
y_{l_i}(s)=[w_{b_1}(s), \cdots, w_{b_{n'_w}}(s),c^T(s)][(r^{agg}_{l_i}(s))^T,e^T_{l_i}]^T, \label{reduced order sys 3}
\end{equation}
where $e_{l_i}\in\mathbb{R}^{n_l}$ is an indicating vector that only the element corresponding to link $l_i$ equals 1, the other elements are zero.  By Eq.~\eqref{disc demand model} and Eq.~\eqref{disc mice demand model}, we have:
\begin{equation}
    \begin{aligned}
        & y_{l_i}(s)= \\
        & [\Bar{w}_{b_1}(k), \cdots, \Bar{w}_{b_{n'_w}}(k),\Bar{c}^T(k)][(r^{agg}_{l_i}(s))^T,e^T_{l_i}]^T+ \\
        & \sum_{j=1}^{n_{\phi}} \phi_j(s\tau;k) [\Bar{\rho}_{j,b_1}(k), \cdots, \Bar{\rho}_{j,b_{n'_w}}(k),\Bar{c}_j^T(k)][(r^{agg}_{l_i}(s))^T, e^T_{l_i}]^T
    \end{aligned}
\end{equation}
Rearrange the above equation we have:
\begin{equation} \label{final equation}
    \begin{aligned}
        y_{l_i}(s)= \begin{bmatrix}
           \Bar{w}_{b_1}(k) \\
           \vdots \\
           \Bar{w}_{b_{n'_w}}(k) \\
           \Bar{c} \\
           \Bar{\rho}_{1,b_1}(k) \\
           \vdots \\
           \Bar{\rho}_{1,b_{n'_w}}(k) \\
           \Bar{c}_1(k) \\
           \vdots \\
           \Bar{\rho}_{n_\phi,b_1}(k) \\
           \vdots \\
           \Bar{\rho}_{n_\phi,b_{n'_w}}(k) \\
           \Bar{c}_{n_\phi}(k) \\
        \end{bmatrix}^T 
        \begin{bmatrix}
            r^{agg}_{l_i}(s) \\
            e_{l_i} \\
            \phi_1(s\tau;k) r^{agg}_{l_i}(s) \\
            \phi_1(s\tau;k) e_{l_i} \\
            \vdots \\
            \phi_{n_\phi}(s\tau;k) r^{agg}_{l_i}(s) \\
            \phi_{n_\phi}(s\tau;k) e_{l_i} \\
        \end{bmatrix}
    \end{aligned}
\end{equation}
The benefits of using the representation Eq.~\eqref{final equation} are two-fold:
\begin{enumerate}
    \item The system order (or the dimension of the inputs) is reduced to $(n'_w+n_l)\times(n_\phi+1)$ in contrast to $(n'_p)\times(n_\phi+1)$ when using Eq.~\eqref{reduced order sys 1}.
    \item Each different routing configuration can provide $n_l$ data points, in contrast to $1$ when using Eq.~\eqref{reduced order sys 1}.
\end{enumerate}
Thus, according to Theorem 1 with $l=1$ and $n=0$, the minimal number of data points required is $\lceil\frac{(n'_w+n_l)\times(n_\phi + 1)}{n_l}\rceil=\lceil\frac{n'_w\times n_\phi + n'_w}{n_l}\rceil+1+n_\phi $. For the Geant topology, if we use two basis functions, the number is only $\lceil\frac{10\times 2 + 10}{36}\rceil+3=4$ we control the top ten elephant flows.

Lastly, we give the following claim:
\\

\noindent\textbf{Claim.} \textit{If $n'_p < n_l \times n'_w$ and $M^{agg}$ has full column rank, then for a given $r^{agg}$, if the equation Eq.~\eqref{M^agg} respect to $r'$ has a solution, then it is unique.}
\\
\noindent This claim means that if we can solve a feasible $r^{agg}$ subject to Eq.~\eqref{M^agg} in a TE problem, then the original routes $r'$ can be uniquely determined.

\section{Data-Enabled Predictive Traffic Engineering}
\label{sec:algorithm}
Now we are ready to present the full algorithm of Data-Enabled Predictive Traffic Engineering (DeeP-TE). 

\subsection{Online Data Sample Collection}
\begin{figure}[htbp]
\centerline{\includegraphics[width=0.45\textwidth,trim={0cm 0cm 0cm 0cm},clip]{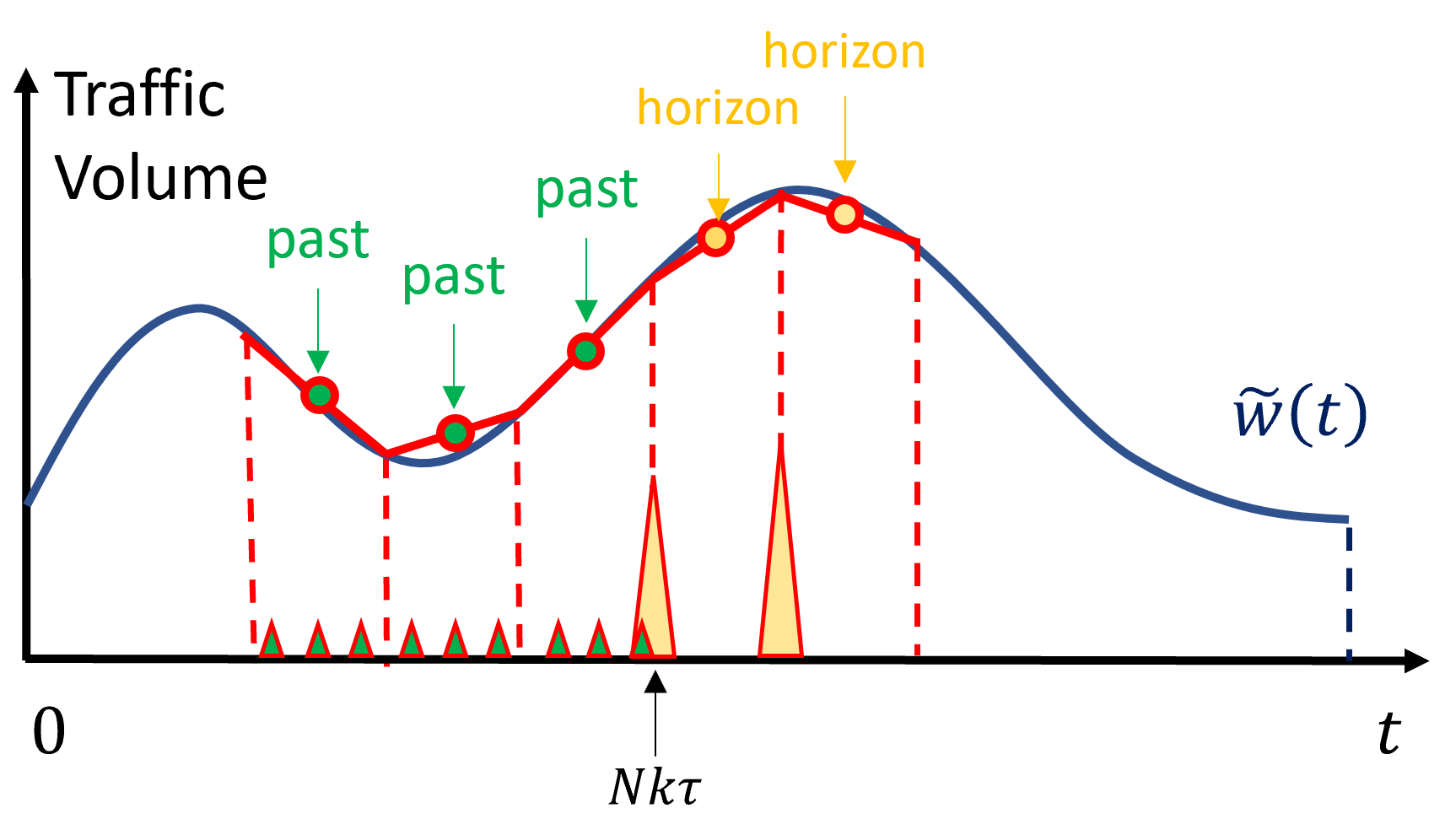}}
\caption{Multi-time Scale Routing Perturbation and Control: Within each control interval, we randomly select demands at each sample interval to apply heuristic routing perturbations (small green spikes), the corresponding link loads will be recorded for data-enabled predictive traffic engineering calculation in the future control intervals (large yellow spikes).}
\label{surrogate}
\end{figure}

As shown in Fig.~\ref{surrogate}, before we compute $r(Nk)$, the old piecewise constant routing configuration $r((k-1)N)$ is sampled for the previous $N-1$ sample intervals. If all the previous $N-1$ sample intervals use the same routing configurations for all the demands, the persistently exciting condition mentioned in Theorem 1 cannot be satisfied. As a result, we need to introduce routing perturbations to a small faction of traffic demands at each sample interval. Any heuristic traffic shifting algorithm can be applied. For example, if a demand $d$ is chosen at one sample interval, due to traffic variations after last routing update $r((k-1)N)$, the delay on the candidate paths of $d$ may no longer be well balanced. We can shift some traffic from a path with longer delay to another path with shorter delay.  
The amount of traffic shift can be randomly chosen from a limited range to control the impact of heuristic routing perturbation while providing enough ``excitation" for DeeP-TE calculation. 

\subsection{Data-Enabled Predictive Traffic Engineering (DeeP-TE)}
Since we would like to minimize the total delay of all links as well as the route changes, we can define the cost function for horizon $h=0,\dots,H-1$ at time step $Nk$ as:
\begin{equation}
J_h(k) = \sum_{i=1}^{n_l} f(y_{l_i}(N(k+h)) / C_{l_i}) + \alpha_1\|\Delta r(N(k+h))\|_1 \label{Cost} 
\end{equation}
where $f$ denotes the function of the piecewise linear approximation of the delay as is introduced in \cite{fortz2000internet}, $\alpha_1$ is some positive constants, $\Delta r(N(k+h))$ denotes $r(N(k+h))-r(N(k+h-1))$. 

Then, as is shown in Fig.~\ref{surrogate}, based on the sampled data, at the beginning of the $k$-th control interval, we can solve the following data-enabled predictive traffic engineering problem:
\begin{subequations} \label{DD-MPC}
\begin{alignat}{2}
& \underset{\textrm{\shortstack{ \tiny $\{\{r,r^{agg},y\}(N(k+h))\}_{h=0}^{H-1}$ \\ \tiny $\left\{\left\{g_{l_i}^{p,h},y_{l_i}^{p,h}(Nk)\right\}_{p=1,h=1}^{L,H-1}\right\}_{l_i \in E}$ }}} {\textbf{minimize}} & & \hspace*{-1.5cm} \sum_{h=0}^{H-1} 0.9^{h} J_h(k) \\ \nonumber
&  + \alpha_2 \sum_{l_i\in E}\left(\sum_{h=0}^{H-1} \sum_{p=1}^{L} \|g_{l_i}^{p,h}(Nk)\|_2\right)  \\
& \text{s.t.} && \nonumber\\ 
& & & \hspace*{-5.6cm} \begin{bmatrix}
r_{l_i}^{agg}(N(k+h)) \\
0_{n'_w\times n_{\phi}} \\
e_{l_i} \\
0_{n_l\times n_{\phi}} \\
y_{l_i}^{p,h}(Nk) 
\end{bmatrix} = 
\begin{bmatrix}
H_{N(k-p-1)+1,N(k-p)-1}^{(r^{agg})}\\ 
(I_{n_\phi}\otimes H_{N(k-p-1)+1,N(k-p)-1}^{(r^{agg})}) S \\
[I_{n_l},\dots,I_{n_l}] \\
(I_{n_\phi}\otimes[I_{n_l},\dots,I_{n_l}]) S \\
H_{N(k-p-1),N(k-p)-1}^{(y^T)} 
\end{bmatrix}g^{p,h}_{l_i}, \label{DD-MPC:sub3} \\
& & & \hspace*{-5.6cm} y(N(k+h))= \sum_{p=1}^{L} X_{p,h} [y^{p,h}_{l_1}(Nk),\dots,y^{p,h}_{l_{n_l}}(Nk)]^T, \label{DD-MPC:sub4} \\ 
& & & \hspace*{-5.6cm} \begin{cases}
r^{agg}_{l_i}(N(k+h))=M_{l_i}r'(N(k+h)), \; l_i \in E, \\
0\preceq r'(N(k+h)) \preceq 1, \\
\sum_{j=1}^{n'_{p,b_d}} r_{[b_d],j}(N(k+h)) = 1, \; d=1,\dots,n'_w;
\end{cases} \label{DD-MPC:sub5}
\end{alignat}
\end{subequations} 
where $\alpha_2$ is a positive constant, and vector $g_{l_i}^{p,h}\in\mathbb{R}^{n_g}$ is the decision variable vector, $n_g=(N-1)\times n_l$ and
\begin{equation} \notag
    S = \begin{bmatrix}
    \text{diag} \left(\left[\phi_1(-\frac{N}{2}+1),\dots,\phi_1(\frac{N}{2}-1)\right]^T \otimes 1_{n_l} \right)\\
    \vdots \\
    \text{diag} \left(\left[\phi_{n_\phi}(-\frac{N}{2}+1),\dots,\phi_{n_\phi}(\frac{N}{2}-1)\right]^T \otimes 1_{n_l} \right)    
    \end{bmatrix}
\end{equation}
in which $1_{n_l}$ ($0_{n_l}$) is an all-one (all-zero) vector of dimension $n_l$, $\otimes$ is the Kronecker product.
For example, using the matrix
\begin{equation} \label{one-basis}
    \text{diag} \left(\left[(-\frac{N}{2}+1),\dots,(\frac{N}{2}-1)\right]^T \otimes 1_{n_l} \right)
\end{equation} 
as the $S$ matrix means that we only adopt one basis function, which is $\phi_{1}(t;k)=t-(k+0.5)\tau_{c}$, as is shown in Fig.~\ref{surrogate}.
Note that, in our case, $N$ is an even integer, and this algorithm can be trivially adapted to the case $N$ is odd.
The last regularization term of the cost function is to enhance the robustness of the algorithm against the data noise induced by the unmodeled traffic demands' fluctuation during the perturbation period \cite{huang2023robust}.

For prediction horizon $0\le h\le H-1$, the constraints Eq.~\eqref{DD-MPC:sub3}-\eqref{DD-MPC:sub4} are data-driven prediction models derived from the reduced-order TE model~\eqref{final equation} and the simple prediction model~\eqref{yp=Br11}, respectively. 
Using $0_{n_l\times n_{\phi}}$ and $0_{n'_w \times n_\phi}$ in the Eq.~\eqref{DD-MPC:sub3} means that the ``dummy'' link load vector $y_{l_i}^{p,h}(Nk)$ is the link load generated by applying $r'(N(k+h))$ to the \textit{mean} value of the past traffic matrix $\Bar{w}(N(k-p))$, as is shown in Fig.~\ref{surrogate}.
%
%
And also note that only perturbation routing data and the corresponding link load data are required. The constraints Eq.~\eqref{DD-MPC:sub5} represent that the fraction of each OD flow routed to some path cannot be negative, and the sum of the fractions of each OD flow routed to all its candidate paths must be one. 

\section{Numerical Experiments}
\label{sec:experiments}
\subsection{General Setup}
\begin{table}
    \centering
    \caption{Topology Information}
	\begin{tabular}{|c|c|c|}
		\hline
		Name & Nodes & Edges \\
		\hline
		france & 21 & 41 \\
		\hline
		geant & 22 & 36 \\
		\hline
		ta1 & 24 & 55 \\
		\hline
	\end{tabular}
	\label{tab:topos}
\end{table}
\textbf{Topology and TM:} 
To evaluate our method, we use three topologies from \cite{orlowski2010sndlib}.
The topologies are filtered so that some closely related nodes are aggregated into one node. The topology information after pre-processing is listed in Table \ref{tab:topos}. The DeeP-TE method does not require measuring or estimating the TMs. If the traffic distribution is known in advance, the traditional methods can first estimate the TM and then optimize the routing.
However, if the traffic distribution is unknown, the estimation step may have large errors, resulting in performance degradation in routing control.
The DeeP-TE method cannot do much better than the traditional methods in the first case, but in the second case, it can avoid performance degradation due to estimation errors.
A typical scenario for the second case is the inter-DC traffic over WAN, where a small group of inter-DC flows accounts for the majority of traffic volume over the background flows.

For each topology, we generate 20 TM time series of the inter-DC traffic over WAN scenario.
The length of each time series is 3 days and the granularity is 5 minutes.
The generation process is based on a widely used traffic generation model \cite{tmgen}.
This model takes into account the spatial and temporal properties of real TMs.
Besides random noise, the generated TMs have sinusoid-like patterns that emulate the diurnal pattern of real network traffic throughout the day.

%
%
We generate the foreground flows and the background flows separately.
The background flows are small flows between all pairs of nodes and account for about 20\% of the total demand volume.
To generate the background flows, we apply the traffic generation model to the whole topology.
The other 80\% demand volume belongs to a small group of large foreground flows, standing for the inter-DC traffic.
The number of foreground flows in each TM series is set as 8.
For the \textit{geant} topology, we first filter out the large flows in the coarse-granularity TMs from \cite{orlowski2010sndlib} whose flow volumes sum to 80\% of all the flows, and the foreground flows are randomly selected among these node pairs.
For the \textit{france} and \textit{ta1} topologies, 7 nodes are first picked as key nodes, and then foreground flows are randomly selected between these key nodes.
We see the selected node pairs as a small sub-topology and apply the traffic generation model again to generate the foreground flows.
As mentioned in Section \ref{sec:5.b}, our method only controls the routing of the foreground flows while the background flows are routed by static routing, e.g., shortest-path. 

After generating the two groups of flows, we add them together to make the complete TM time series.
The last step for generation is to scale the TM time series to make proper network load levels.
On one hand, if the network load level is too low, the delay remains low for all routing strategies. On the other hand, if the network load level is too high, no routing configuration could avoid congestion. Either case would make all methods perform similarly.
The scaling factors are determined for each TM series so that the optimal routing would have the average link utilization around 30\% to 40\% or the maximum link utilization close to 1.


The control interval is set as 30 minutes, while the sample interval is set as 5 minutes. 
This means that our DeeP-TE algorithm would change the routing configuration decision every 30 minutes, and inject different perturbations as well as collect link load data every 5 minutes.
Each flow can be split arbitrarily into its top four shortest paths.

\textbf{Prediction Model:}
We use the last 3 control intervals' traffic data to predict the next 2 control intervals' traffic data; that is, we set $L=3$ and $H=2$.
To compute the prediction model $X$ mentioned in Section \ref{sec:MPC} for each time series, we use the traffic data $\{w^d(1),\cdots,w^d(D)\}$ of the first two days, averaged in each control interval to generate the following data matrices:
\begin{subequations}
\begin{alignat}{2}
& W^{\text{history}} := \nonumber \\ \nonumber
    &[\text{vec}(H^{(w^d)}_{1,D-H-L+1}), \text{vec}(H^{(w^d)}_{2,D-H-L+2}), \cdots, \text{vec}(H^{(w^d)}_{L,D-H})] \\ \nonumber 
& W^{\text{future}} := \\ \nonumber
    &[\text{vec}(H^{(w^d)}_{L+H,D}), \text{vec}(H^{(w^d)}_{L+H-1,D-1}), \cdots, \text{vec}(H^{(w^d)}_{L+1,D-H+1})], \nonumber
\end{alignat}
\end{subequations}
 where $\text{vec}(x)$ means putting $x$ into vector form.
Then we can define the optimal $X^*$ by the following optimization problem:
\begin{equation}
\begin{aligned}
X^* := \underset{X}{\text{argmin}} \|W^{\text{future}} - W^{\text{history}}X\|_2 \nonumber
\end{aligned}
\end{equation}

Note that even though the prediction model is obtained by fitting the historical TM data, it is only used to predict future link loads.
Our algorithm does not explicitly use the TM history or predict the future TM, but directly controls the future routing configurations.

\textbf{Perturbation:} The random perturbations to the traffic splitting ratios of the elephant flows are generated from a Dirichlet distribution for $5\%$ of elephant flows in each sample interval between two adjacent control inputs.
Suppose the current time step is $s$ and the splitting ratio of elephant flow $i$ is $r_i(s)$. To generate the perturbed splitting ratio in the next step $r_i(s+1)$, we draw a sample $\hat{r}$ from the flat Dirichlet distribution of the same dimension as $r_i(s)$, and set $r_i(s+1) = 0.95 r_i(s) + 0.05 \hat{r}$.

Following the control interval and sample interval settings, the number of data points collected in each control interval is 6, exceeding the minimal requirement of 4 data points calculated as in section \ref{sec:system-order-c}.
However, since the perturbations are generated randomly, the optimization problem (\ref{DD-MPC}) could sometimes be infeasible.
In this situation, the routing configurations stay unchanged as in the last control interval.
We count the number of times when this situation happens, and it is only about 2\% of the total decision times.

\textbf{Baseline Methods:} We compare our proposed algorithm with the following algorithms:
\begin{itemize}
    \item \textbf{Tomogravity MPLS (TG)} \cite{roughan2003traffic}: This method utilizes the TM estimated by the tomogravity algorithm at each interval to set the routing configurations of each elephant flow using the link-path formulation to minimize the total delay.
    The control interval is 5 minutes or 30 minutes.
    \item \textbf{Constant Routes (CONST)}: This method uses the mean values of TM data of the first two days to set the static optimal routing configurations of elephant flows using the link-path formulation to minimize the total delay and keep such routes unchanged all the time.
    \item \textbf{Optimal Routes (OPT)}: This method uses the real value of the TM at each time step, acting like an oracle for the best delay performance. It optimizes the routes for \textit{all} flows to minimize the delay but does not consider the route changes between consecutive steps.
\end{itemize}

\textbf{Metrics:} We look at two metrics:
\begin{itemize}
    \item \textbf{Performance Ratio (PR)}: We denote the optimal link loads at time step $s$ obtained by OPT model as $y^{Opt}(s)$ and define the following performance ratio (PR) as our control performance metric:
        \begin{equation} \label{PR}
        PR(s) := \frac{\sum_{i=1}^{n_l} f\left(y_{l_i} (s)/ C_{l_i}\right)}{\sum_{i=1}^{n_l} f\left(y^{Opt}_{l_i} (s)/ C_{l_i}\right)} \nonumber
        \end{equation}
    PR is always greater than or equal to 1.
    \item \textbf{Route Change (RC)}: This metric is about the difference in routing configurations between two consecutive time steps:
        \begin{equation} \label{RC}
        RC(s) := \|r(s + 1) - r(s)\|_1 \nonumber
        \end{equation}
\end{itemize}

The metric value for the whole TM time series is the average of that at each time step.
Both metrics are better when smaller.

\subsection{Effect of Basis Functions}

As mentioned in Section \ref{sec:5.a}, we assume that we can use basis functions to model the traffic dynamics within a control interval.
The basis functions are known while their coefficients are missing.
However, the perfect knowledge of basis functions could be infeasible in practice.
The error in the traffic dynamic model could lead to a potential performance gap.
Using more basis functions would reduce the error and narrow the gap, but meanwhile make the problem (\ref{DD-MPC}) too large to solve.
As a result, the selection of basis functions is important to tradeoff between routing performance and solution complexity. 

We show the importance of choosing appropriate basis functions through a small-scale experiment.
We pick one TM time series, and then process it to be constant or linear within each control interval by doing a least-square approximation.
Finally, we add back some noise to these three TM time series while keeping the level of noise unchanged.
The generated TMs are called original-sinusoid, piecewise-constant, and piecewise-linear respectively.
For the DeeP-TE method, we use only the linear basis function, as in Eq. (\ref{one-basis}).

\begin{figure*}[htbp]
\centering
\begin{minipage}{0.33\textwidth}
\centerline{\includegraphics[width=2in,height=1.5in]{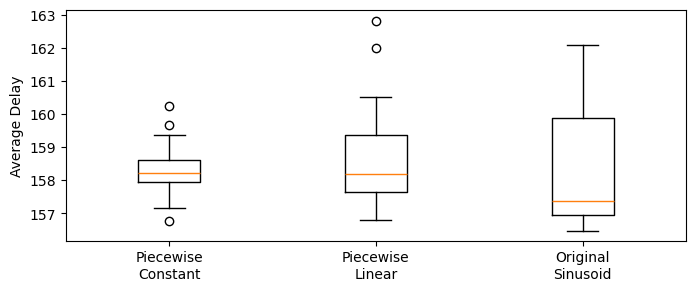}}
\vspace{-0.1in}
\caption{The performances of the DeeP-TE models with only linear basis functions on piece-wise constant, linear and sinusoid TM patterns.}
\label{fig:basis}
\end{minipage}
\begin{minipage}{0.66\textwidth}
  \begin{center}
    \subfloat[]{\includegraphics[width=2in,height=1.5in]{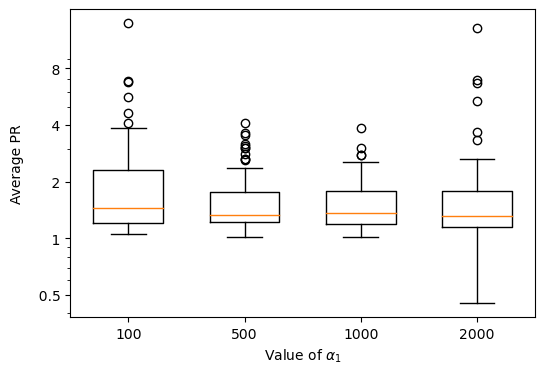}
    \label{fig:alpha1-PR}}
    \hfil
    \subfloat[]{\includegraphics[width=2in,height=1.5in]{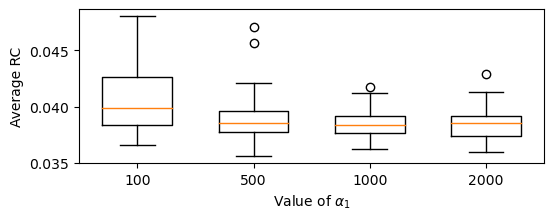}
    \label{fig:alpha1-RC}}
    \caption{Impact of Penalty Weight $\alpha_1$: (a) average PR and (b) average RC.}
    \label{fig:alpha1}
    \end{center}
\end{minipage}
\end{figure*}
We run the DeeP-TE method multiple times, and Fig. \ref{fig:basis} shows the average delay results as a box plot.
Using only the linear basis function, the DeeP-TE method encounters a larger variation in delay on the original-sinusoid TMs than on the other two groups of TMs. 
The large variations here come from errors in traffic dynamics.
The piecewise-constant and piecewise-linear dynamics could fit into the linear basis functions, but the original-sinusoid dynamics cannot be completely captured by the linear basis function, thus leading to larger variations.
The difference in delay is not large, because the sinusoid-shape TM has a long cycle of several hours, and the traffic dynamics in most control intervals of 30 minutes is almost a straight line.
However, if the TM is more dynamic or the link capacities are limited, the larger variations could lead to unfeasible optimization steps and further degradation in performance.
Therefore, we use two basis functions in the DeeP-TE method in the following experiments, one linear and the other quadratic, since they can provide enough flexibility in approximating the curves.

\subsection{Effect of Penalty Weight $\alpha_1$}

We vary $\alpha_1$ to show the tradeoff between mitigating the route changes and the total delays. The $\alpha_1$ values are chosen from $\{100, 500, 1000, 2000\}$. Ideally, given a perfect traffic prediction and variation model, the smallest $\alpha_1$ should achieve the best performance, but incur the largest route changes. 

We calculate the average PR and RC for each TM time series, and the results are shown in Fig.~\ref{fig:alpha1}.
As $\alpha_1$ increases, the RC generally decreases as expected, but the PR also decreases.
This is because our traffic models are not perfect.
Some penalties on the route changes can help to stabilize network routing when traffic variations are not accurately modeled by the linear prediction model and basis functions.
Since there are many outliers in Fig. \ref{fig:alpha1-PR} when $\alpha_1$ is 500 or 2000, we choose $\alpha_1 = 1000$ for comparison study with other baselines.



\subsection{Comparisons with Baselines}
\begin{figure}[htbp]
\centerline{\includegraphics[width=0.4\textwidth]{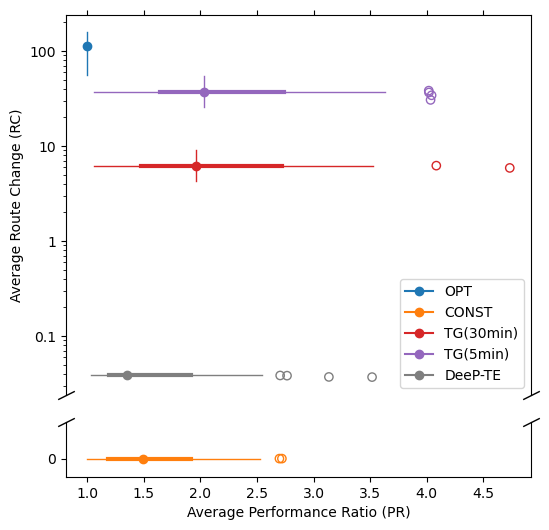}}
\vspace{-0.1in}
\caption{The performances of different algorithms on the entire dataset are shown.  The X-axis is the average PR, and the Y-axis is the average RC in the log scale. Each cross indicates the distribution of data points for each algorithm. The vertical line denotes the range of RC, while the horizontal line follows the box plot method, with the ends of the thick line marking the first and third quartiles. Outliers are indicated by hollow dots. Better performance is represented by points closer to the bottom left corner.}
\label{fig:main}
\end{figure}

We run different algorithms over the 60 generated TM time series, and the results are shown in Fig. \ref{fig:main}.
The OPT method always performs the best because it has knowledge of the real TM data.
The route changes are not in the objective function to be optimized at each step, resulting in high RC values.
On the other extreme, the CONST method always has the RC equal to 0 since it does not change the routes.
The performances of the two TG methods are bad because the gravity TM estimation model they use is not a good fit in our scenario.
The estimated TM is usually biased, leading to performance degradation.
The low rerouting frequency of the TG(30min) model helps to avoid unnecessary adjustments, thus leading to lower RC and slightly better PR than those of the TG(5min) model.
Our DeeP-TE method can achieve good performance while keeping the routes stable.
It has similar quartiles as the CONST method, but the median point is closer to the left, meaning that the distribution of points is more dense to the left. 
The RC values of our method are much lower than those of the two TG or the OPT methods, because our method only adjusts the base routing configurations when necessary.

\begin{figure*}[!t]
\centerline{\includegraphics[width=1.0\textwidth]{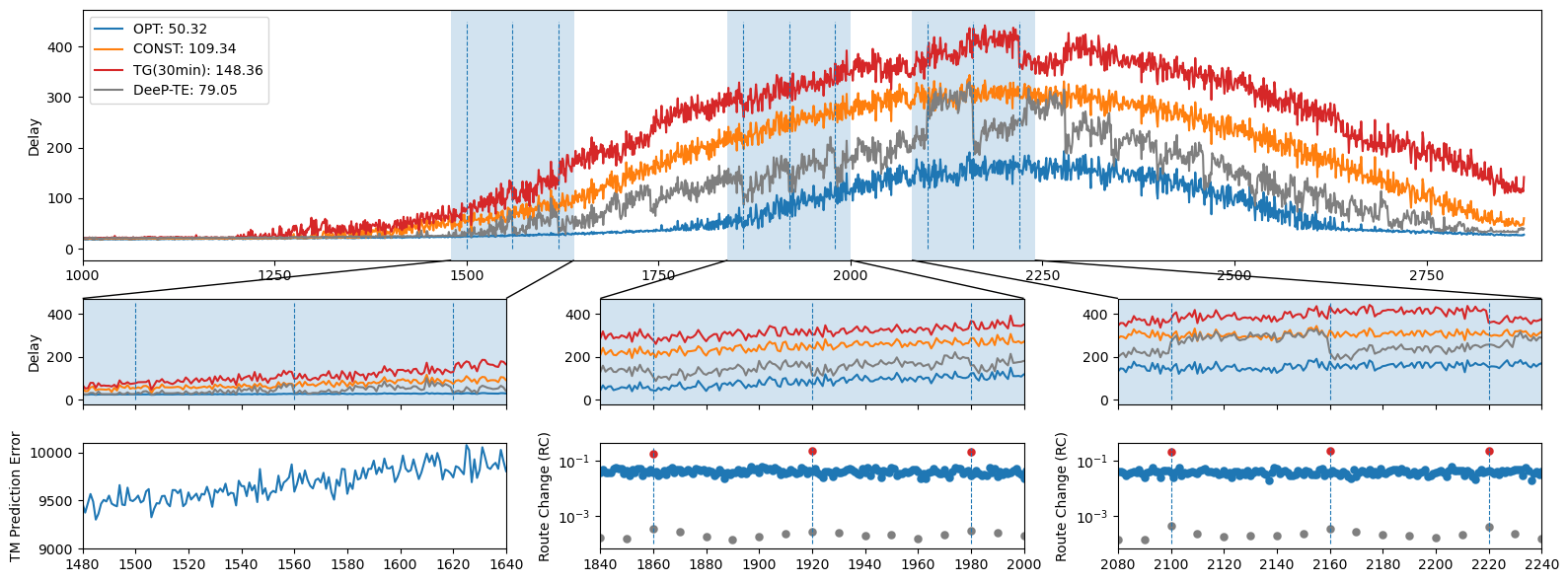}}
\vspace{-0.1in}
\caption{The performances of different methods on the one typical TM time series are shown. The upper figure shows the total delay over one day. The lower three groups of figures are the detailed looks within three sub-periods, each group shows the delay and the TM prediction error of the TG method or the route changes. The TM prediction error is the L2-norm between the real and the estimated vectorized TMs. Each control interval is 30 minutes and the beginning is marked with vertical dotted blue lines. The color of the dots in the RC figures follows the legend in the upper figure.}
\label{fig:zoom}
\end{figure*}

To understand further how our method works, let's zoom into a typical TM time series to see the details.
We choose one time series on the topology \textit{geant}, and the results are shown in Fig. \ref{fig:zoom}.
In the first eight hours, all methods perform similarly, so that period is omitted.
Looking at the upper figure of the total delay, we can see that the DeeP-TE method has the closest delay performance to the OPT method.

To maintain optimal performance, the OPT method adjusts the routing configurations at each time step, resulting in consistently high RC values over time as shown in the RC figures in the last row.
The TG method changes the routing configuration every 30 minutes based on its TM estimation.
However, since the tomogravity TM model is unaware of the inter-DC traffic scenario, the TM estimation incurs a large error.
The error keeps increasing during the first slice as shown in the bottom-left figure, which is caused by the increasing traffic volumes in the period.
Even though the TG method makes a big change to the routing configurations at the beginning of each interval, the delay performance does not benefit that much.
The CONST method does not change the routing configurations.
It has a high delay when the traffic volume increases, yet slightly better than the TG method, indicating the big impact of biased TM estimation.
In contrast to the frequent adjustments by the OPT method, our DeeP-TE method avoids unnecessary routing configuration changes.
The RC values at the times when the perturbations are injected are usually less than those at the beginning of control intervals.
In the first two slices, the delay performance of the DeeP-TE method is very close to that of the OPT method in the first 5 minutes of each control interval.
This is because there is no perturbation injected in the first 5 minutes and the method could find a solution so close to the optimum.
The injected perturbations lead to the later small gaps, while they help to solve the optimal routing configurations in the following intervals.

The last slice shows the drawbacks of our DeeP-TE method.
To bypass the TM estimation step, our method relies on the randomly generated perturbations to infer the optimal way to adjust the future routing configurations.
However, the link load measurements would simultaneously be affected by the injected perturbations and traffic volume dynamics.
Our method cannot perfectly distinguish the cause of the link load changes, resulting in some mistaken adjustments and sub-optimal solutions.
For example, in the control interval starting from timestamp 2100, the delay performance of our method is not close to the optimum.
The gaps are usually resolved very soon, as shown in the next interval starting from timestamp 2160.
In later intervals and in other traces, this phenomenon occurs several times.
We will find better ways to generate the perturbations to solve this issue in future work.

\section{Conclusions}
\label{sec:conclusion}
Using the behavioral approach to system theory, we developed a novel data-enabled predictive control-based adaptive routing algorithm that generates routing updates based on data samples of historical routing configurations and the associated link rates, without direct TM measurements/estimations. Through analysis and experiments, we demonstrated that the number of data samples needed is well manageable in practice by focusing on dominating elephant flows and aggregating routing configurations using network topology information. In experiments driven by real network topologies and TMs, our algorithm achieved close-to-optimal control effectiveness with significantly lower routing variations than the baseline methods. Our first attempt to apply data-enabled predictive control to network traffic engineering is encouraging. In our future work, we will further analyze the robustness of our algorithm against the magnitude of traffic variations and the error of the dynamic traffic model. We will also develop a hybrid solution where a portion of TM entries can be directly measured or accurately estimated.


\bibliography{ref}

\end{document}